\input{aipcheck}

\newcommand{\nn}{\nonumber}
\newcommand{\beq}{\begin{equation}}
\newcommand{\eeq}{\end{equation}}
\newcommand{\bqa}{\begin{eqnarray}}
\newcommand{\eqa}{\end{eqnarray}}
\newcommand{\ben}{\begin{eqnarray*}}
\newcommand{\een}{\end{eqnarray*}}

\documentclass[
   ,final            
  ]
  {aipproc}

\layoutstyle{6x9}

\begin{document}

\title{Model-independent Study on Magnetic Dipole Transition
in Heavy Quarkonium}


\author{Yu Jia}{
  address={INFN and Dipartimento di Fisica dell' Universita di Milano, via Celoria 16,
20133 Milano, Italy},
  email={frank.mittelbach@latex-project.org},
  thanks={This work was commissioned by the AIP}
}



\begin{abstract}

Some new results on  magnetic dipole ($M1$) transitions in heavy quarkonium from 
nonrelativistic effective field theories of QCD are briefly reported.
This model-independent approach not only facilitates a systematic and lucid way to 
investigate the relativistic corrections, it also clarifies some inconsistent
treatment in previous potential model approach. The impact of our
formalism on $J/\psi\to \eta_c\gamma$, $\Upsilon(\Upsilon^\prime)\to \eta_b\gamma$ and 
$h_c\to\chi_{c0}\gamma$ are discussed.

\end{abstract}

\date{\today}

\maketitle


Radiative transitions in heavy quarkonium are of considerable experimental and theoretical
interest~\cite{Brambilla:2004wf}. 
On the theory side, it provides us with further insight on the dynamics of
quarkonium in addition to the knowledge we have gleaned from the spectra
of $c\bar c$ and $b \bar b$ families.
 
Being an old subject, radiative transitions have been extensively studied 
in phenomenological models, notably the potential model approach~\cite{Feinberg:1975hk}.  
It is certainly desirable to study them from a model-independent perspective.
In this talk I will report such a study based on the effective-field-theory (EFT)  
approach~\cite{commonpaper}.  
Since $M1$ transition is theoretically cleaner and more interesting than $E1$ transition, 
I will focus on the former case, despite the fact that the latter is 
observed more copiously in nature.

In the nonrelativistic limit, the 
M1 transition rate between two $S$-wave onia takes a particularly simple form:
\bqa
\Gamma[n\:{^3S_1}\to n^\prime\:{^1S_0}+ \gamma] &=&  
{ 4  \alpha_{\rm em} e_Q^2 \over 3 m^2}
(1+\kappa_Q)^2 \, k_\gamma^3\, \left|\int dr r^2 R_{n^\prime 0} R_{n0} \right|^2 \,,
\label{M1:NRlimit}
\eqa
where $e_Q$ and $\kappa_Q$ are 
the electric charge and anomalous magnetic moment of the heavy quark, 
and $R_{nl}(r)$ is the radial Schr\"{o}dinger wave functions.
Since the leading dipole operator $c_F e e_Q/2m \, \sigma\cdot {\bf B}^{\rm em}$ 
(where $c_F=1+\kappa_Q$) 
only flips the spin and doesn't act on the spatial degrees of freedom, 
orthogonality of radial wave functions guarantees that: 
in the \emph{allowed} transition ($n=n^\prime$), the overlap integral equals 1;
in the \emph{hindered} transition ($n \neq n^\prime$), the overlap integral vanishes.

It is well known that \eqref{M1:NRlimit} overpredicts the observed
$J/\psi\to \eta_c \gamma$ transition rate by a factor of $2\sim 3$,
with a normal input of $m_c$ and $\kappa_c$ from perturbative one-loop matching.
This clearly indicates large relativistic corrections to \eqref{M1:NRlimit}, 
as is usually confronted in charmonium system. It has also been speculated 
that low-energy fluctuations may generate a large negative $\kappa_c$, 
so the discrepancy can be reduced.

The ${\cal O}(v^2)$ corrections to \eqref{M1:NRlimit}  has been available for a long time
from potential model approach~\cite{Brambilla:2004wf, Feinberg:1975hk}:
\bqa
\Gamma[n\:S\to n^\prime\:S +\gamma] &=&  {1\over 2J_n+1}
{ 4  \alpha_{\rm em} e_Q^2 \over  m^2} \,k_\gamma^3\, \left|\, \sum_i^5 I_i \, \right|^2 \,,
\label{M1:v2:correction}
\eqa
where $2 J_n+1$ counts number of polarizations of the parent onium, and
\bqa
I_1 &=& \left \langle n^\prime 0 \left| (1+\kappa_Q) \left( 1-{k_\gamma^2 r^2\over 24} \right) +
( 1+ 2 \kappa_Q) {k_\gamma\over 4 m}  \right|n 0\right \rangle  \,,
\\
I_2 &=& - \left \langle n^\prime 0\left| (1+\kappa_Q) {{\bf p}^2 \over 2 m^2} +
{{\bf p}^2\over 3 m^2}  \right| n 0\right \rangle  \,,
\nn \\
I_3 &=&  \left \langle n^\prime 0\left| {\kappa_Q \over 6 m} r V_0^\prime  \right|n 0\right\rangle  \,,
\nn  \\
 I_4 &= & \pm {4\over E^{(0)}_{n0}-E^{(0)}_{n\prime 0}}  
  \left\langle n^\prime 0 \left| (1+\kappa_Q) {V_{\rm ss} \over m^2} \right|n0 \right\rangle \,,
\nn \\ 
I_5 &=& \left \langle  n^\prime 0\left| -{\eta \over m}  V_{\rm S} \right| n 0\right \rangle \, ,
\nn
\eqa
where 
$V_0$ stands for the static potential, the ``+/--" sign in $I_4$  is associated with 
${^3S_1}\to {^1S_0}$ and ${^1S_0}\to {^3S_1}$, respectively.
Notice that $I_4$ accounts for the first-order correction to 
the wave function due to spin-spin potential (other higher dimensional {\it local} potentials
cease to contribute in $S$-wave transition), thus is only present in hindered transition.

$I_5$ is a prediction specific to the popular assumption in potential models, where one usually decomposes
the confining potential  into a Lorentz scalar and a vector
part, $V_{\rm conf} = \eta V_{\rm S} + (1-\eta) V_{\rm V}$. 
This term constitutes the major uncertainty in the potential model predictions, where contradictory
claims often appear in the literature~\cite{Feinberg:1975hk}.

Presence of a  hierarchy of scales in quarkonium,
$m$, $mv$, $mv^2$, $\Lambda_{\rm QCD}$,  makes the EFT approach an ideal tool to analyze this 
transition process. 
One first descends from QCD to NRQCD by integrating out
hard modes ($\sim m$ )~\cite{Bodwin:1994jh}, then descends from NRQCD to potential NRQCD (pNRQCD) by  further integrating out soft
modes ($\sim m v$ )~\cite{Pineda:1997bj}. 
As a result, the inter-quark potentials appear as Wilson coefficients in pNRQCD, 
and the only dynamical degrees of freedom of pNRQCD are ultrasoft modes ($\sim m v^2$ )
(for convenience, one can also incorporate the radiated ultrasoft photon into pNRQCD).

A primary task of the EFT approach is to validate/invalidate 
\eqref{M1:v2:correction}. The real strength of the EFT approach is, however, that it  can further                
answer the following questions that are beyond the scope of potential models:
\begin{itemize}
\item
 Is  it possible that a large correction to $\kappa_Q$ due to soft modes 
 arises when one descends from NRQCD to pNRQCD? 
\item
 Is it possible to reproduce $I_5$ in pNRQCD? If so, how to interpret it? 
\item
Potential model focus exclusively on the $Q\overline{Q}$ Fock-state 
(with an exception of coupled-channel effects
which may not be relevant as long as one excludes those states close to the open-flavor threshold). 
pNRQCD allows one to  include ultrasoft gluons as dynamical degrees of freedom.
Therefore, one naturally asks for the possibility of 
large nonperturbative contribution arising from higher-Fock states  $|Q \overline{Q} g\rangle$.
This effect is usually referred to as \emph{color-octet effect}.
\end{itemize}

In fact, the answers to these three questions are all \texttt{negative}~\cite{commonpaper}, 
on which we will elaborate shortly. 
It turns out that the pNRQCD formalism is able to
justify  \eqref{M1:v2:correction}  except $I_5$.  In the
so called \emph{weak-coupling regime} (when $mv\gg \Lambda_{\rm QCD}$), this formula is complete;
in the so-called \emph{strong-coupling regime} (when $mv\sim \Lambda_{\rm QCD}$), however,
\eqref{M1:v2:correction} is incomplete and further terms are needed.

Before going on to explanations, it is useful to first sketch the derivation of ${\cal O}(v^2)$
corrections in the framework of pNRQCD. One obvious source is from the contribution of
higher dimensional $M1$ operators.  
These operators can be identified most conveniently by promoting the 
color gauge group of NRQCD Lagrangian to 
a larger gauge group $SU_c(3)\times U_{\rm em}(1)$. Explicitly, 
the relevant magnetic operators up to ${\cal O}(1/m^3)$ read~\cite{Manohar:1997qy}:
\bqa
{\cal L}_{\rm NR} &=& \psi^\dagger \left( i D_0 + {{\bf D^2} \over 2m} \right) \psi
+ {c_F e e_Q\over 2 m} \psi^\dagger {\bf \sigma} \cdot {\bf B}^{\rm em} \psi
+ {i\,c_S e e_Q \over 8 m^2} \psi^\dagger {\bf \sigma} 
\cdot [\nabla\times,{\bf E}^{\rm em}]\psi  
\nn \\ 
&+ & {c_{W1} e e_Q\over 8 m^3} \psi^\dagger \{ \nabla^2, {\bf \sigma} \cdot {\bf B}^{\rm em} \}\psi
-
 {c_{W2} e e_Q\over 4 m^3} \psi^\dagger 
 \nabla_i {\bf \sigma} \cdot {\bf B}^{\rm em} \nabla_i \psi
 \nn \\
&-& {c_{p^\prime p} e e_Q\over 8 m^3} \left[
\nabla \psi^\dagger \cdot {\bf \sigma} \, {\bf B}^{\rm em}\cdot \nabla\psi 
+ \nabla\psi^\dagger \cdot {\bf B}^{\rm em} {\bf \sigma}\cdot \nabla \psi 
\right] + ({\rm \psi \rightarrow i \sigma ^2 \chi^*}) \,,
\label{NRGT:SU3xU1}
\eqa
where $D_\mu\equiv \partial_\mu +i g T^a A^a_\mu + ie e_Q A^{\rm em}_\mu$.
Various Wilson coefficients can be computed through  
perturbative matching at the hard scale.
For example, at one loop accuracy, $\kappa_Q =C_F \alpha_s/ 2\pi$,
is about a few percent for charm and bottom.
It is well known that some of the Wilson coefficients are 
related with each other because of reparameterization invariance (RPI), or essentially
Poincare invariance, 
notably
$c_S = 2 c_F -1$, $c_{W2} = c_{W_1} -1$ and $c_{p^\prime p} = c_F -1$~\cite{Manohar:1997qy}. 
When inherited into pNRQCD, one make replacement $\nabla\to i {\bf p}$ in these operators,  
and these RPI relations can be utilized to condense the expressions.

There are other sources of ${\cal O}(v^2)$ corrections,
for instance, multipole expansion of the photon field. 
One interesting contribution, 
first pointed out by Grotch and Sebastian~\cite{Feinberg:1975hk},
is  the Lorentz boost effect due to the final-state recoil. 
Since the wave function of a moving $S$-wave state has a non-vanishing overlap 
with a $P$-wave, spin-flipped state at rest, the $M1$ transition can be effectively 
realized by a ``E1" transition from the parent to this small component. 
Some subtlety arises in this effect, namely the recoil correction depends on which  ``E1" operator
one uses, {\it i.e.}, $2 e e_Q/m\, {\bf p}\cdot {\bf A}^{\rm em}$ or 
$e e_Q{\bf r}\cdot {\bf E}^{\rm em}$, which are connected by a redefinition of pNRQCD field.
The solution to this problem is as following.  The matching from NRQCD generates 
a  new $M1$ operator at ${\cal O}(1/m^2)$, which is intimately related to the 
{\it non-local} (depending on the center-of-mass momentum) 
spin-orbit potential, with a coefficient proportional to 
$V_{LS}^{\rm CM} = - V_0^\prime / 8 r$  (Gromes relation).  
The contribution of this operator also depends on the field redefinition.
However, the sum of these two corrections is convention-independent, 
thus comprises a meaningful relativistic recoil effect.

We are now in a position in discussing general matching of $M1$ operators from
NRQCD to pNRQCD. At ${\cal O}(1)$, those operators in (\ref{NRGT:SU3xU1}) 
are trivially inherited to pNRQCD. In general, after integrating out soft modes, 
new Wilson coefficients will depend on inter-quark separation $r$.

Obviously the correction to $c_F$ can be interpreted as a multiplicative 
${\cal O}(1/m^0)$ matching coefficient of the leading $M1$ operator.
Dimensional considerations require that this correction is function of $\log(r)$.
We emphasize in passing  that it is inappropriate to attribute this new coefficient to 
the magnetic moment of an individual quark, since it
really arises from entangled contributions from  both quarks.

\begin{figure}[bt]
  \resizebox{20pc}{!}{\includegraphics{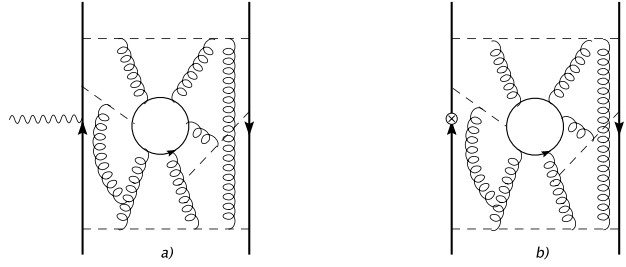}}
\caption{ Typical NRQCD diagrams responsible for matching of
the dipole operator at ${\cal O}(1/m^0)$.
The left diagram represents the vertex correction, and the right
one contributes to the wave-function renormalization factor $Z^{\rm s}(r)$ due to
soft modes,
where the cross-cap implies insertion of  a unit operator.}
\label{leading:M1:op:matching}
\end{figure}

The simple answer, no corrections at all, is basically nothing more than 
the heavy quark spin symmetry and 
that the $\sigma\cdot {\bf B}^{\rm em}$ operator behaves like a unit operator 
in spatial Fock space.
These facts are independent of whether the matching is performed perturbatively 
as in weak-coupling regime,
or nonperturbatively, as in strong-coupling regime. 
In an arbitrary NRQCD diagram depicting transition, as shown in 
Fig.~\ref{leading:M1:op:matching}a),
we only need to consider the case where the external ultrasoft photon is attached 
to one of the heavy quark lines.
(photon attached to internal light quark loops can be neglected 
when summing over electric charges of three light flavors).
Soft gluons attached to heavy quark lines must be longitudinal to avoid $1/m$ suppression. 
Because all the propagators and vertexes 
are spin-independent except the $M1$ vertex, such a diagram can factorize into 
the leading $M1$ operator times minus the wave function renormalization constant $\delta Z_s(r)$,
which can be extracted from the same diagram with the electromagnetic vertex
replaced by a unit operator insertion,
as shown in  Fig.~\ref{leading:M1:op:matching} b).
Therefore, there is no net contribution to this matching coefficient from soft modes.

The matching  at ${\cal O}(1/m)$ is slightly more complicated. In a NRQCD diagram
with a insertion of the leading dipole opeartor, a transverse gluon is allowed
to end on the heavy quark line with a spin-independent ${\bf p}\cdot {\bf A}$ vertex 
(ending on  a $\sigma\cdot {\bf B}$ vertex doesn't contribute).
Repeating the previous argument,  one doesn't obtain any new $M1$ operator from these sources.
However, there are a new class of diagrams, in which electromagnetic coupling is embedded in the
covariant derivative $\bf D$, inducing an effective magnetic operator at ${\cal O}(1/m)$.
This is just the one accompanying with the nonlocal $LS$-potential as mentioned before.
It is important to note that there is no room at this order for the operator of the form
$V_{\rm S}/m^2 \sigma\cdot {\bf B}^{\rm em}$ to arise.
Therefore one is forced to regard $I_5$  in (\ref{M1:v2:correction}) 
as an artifact of potential model, 
which cannot bear any physical significance.

The matching at ${\cal O}(1/m^2)$  becomes much more cumbersome.
We will not dwell on the explicit expressions for these case.
However, it should be kept in mind that they may be neglected 
in the weak-coupling regime because of suppressions by higher powers of
$\alpha_s\sim v$. In the strong-coupling regime, since $\alpha_s\sim 1$, 
these operators might be as important as other ${\cal O}(v^2)$  corrections. 
These corrections involve some unknown nonperturbative Wilson loop amplitudes,
so that predictive power is unavoidably damaged.

\begin{figure}[bt]
  \resizebox{25pc}{!}{\includegraphics{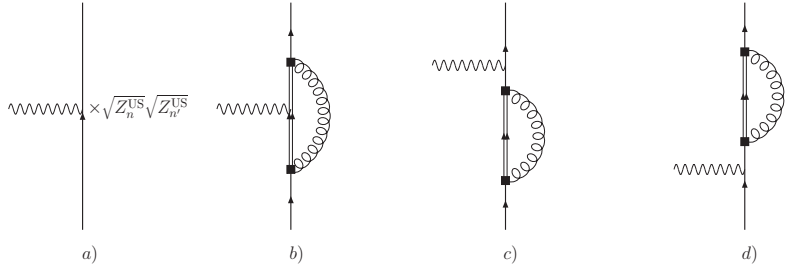}}
\caption{pNRQCD diagrams for color-octet contribution to  radiative transition. 
Single and double lines correspond to the singlet and octet fields. 
The singlet-octet-gluon vertex is of chromo-E1 type.} 
\label{co:pNRQCD:M1:diagrams}   
\end{figure}

The color-octet effect to radiative transitions was first envisaged by Voloshin long time ago,
but without a detailed study~\cite{Voloshin:1978hc}. 
pNRQCD allows a systematic treatment of this effect. 
The corresponding diagrams are shown in Fig.~\ref{co:pNRQCD:M1:diagrams}.  
It is not surprising that the color-octet effect is nearly absent in $M1$ transition,
because of the exactly same reason as before. It is interesting to note this bears some resemblance as
absence of leading nonperturbative correction to $\overline{B}\to D^*$ form factor at the zero recoil, known as
Luke's theorem~\cite{Luke:1990eg}.
Here we only give a heuristic argument based on the quantum-mechanical 
perturbation theory. 
Treating the chromo-E1 interaction as perturbation
up to second order,
one can schematically  express a ``true'' quarkonium state as superpositions of the following Fock components (stripping off
spin d.o.g.):
\bqa
   |N \rangle = \sqrt{Z_n^{\rm us}}| Q\overline{Q}_1 (n) \rangle  + |Q \overline{Q}_8 g\rangle + 
 \sum_{m\neq n} | Q\overline{Q}_1 (m) \rangle  \cdots\, ,
\label{Fock:expansion:co:2ndorder} 
\eqa 
where $| Q\overline{Q}_1 (n) \rangle$ is the unperturbed state, and  $Z_n^{\rm us}$ is the
wave function renormalization factor for state $n$ due to ultrasoft gluons.
Spin-independence of chromo-E1 interaction,  
plus $M1$ operator being a unit operator in coordinate space, imply that the full $M1$ transition
amplitude is nothing but the inner product $\langle N^\prime | N \rangle$. Since orthogonality condition 
is preserved in perturbed states, it is equal to $\langle n^\prime | n \rangle$, 
the color-octet contribution thus vanishes. We emphasize that sizable color-octet effect might arise in $E1$ transition,
since the $E1$ operator is no longer a unit operator in coordinate Fock space.

Finally we turn to the phenomenological implication of \eqref{M1:v2:correction},
with the understanding that we restrict only to the weakly-coupled system for
consistency and the unphysical $I_5$ term has been dropped.
Thus far, the only observed $M1$ transitions are $J/\psi (\psi^\prime ) \to\eta_c\gamma$, 
and upper bounds on $\Upsilon^\prime (\Upsilon^{\prime\prime}) \to\eta_b\gamma$ 
have recently been set by CLEO~\cite{Artuso:2004fp}. 
Empirically, $\Upsilon(1S)$ and $\eta_b$  are believed to lie in the weak-coupling regime, 
whereas $J/\psi$, $\eta_c$  fit in this regime to a less extent. 
$\Upsilon^\prime$ and $\Upsilon^{\prime\prime}$ are usually regarded as 
strongly-coupled system. 
As for $\psi^\prime$ and $\eta_c^\prime$, they are too close to the open-flavor threshold and 
cannot be correctly described by current formulation of pNRQCD, 
therefore we exclude them in our analysis.

\begin{figure}[bt]
  \resizebox{18pc}{!}{\includegraphics{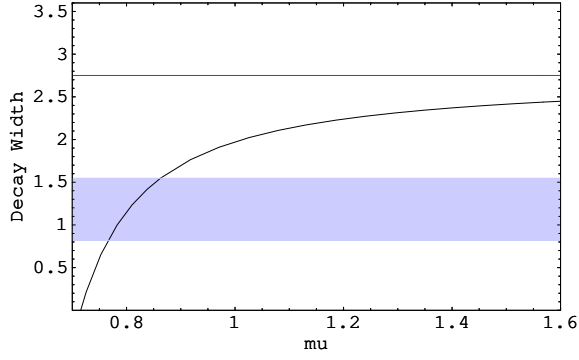}}
\caption{ $\Gamma[J/\psi\to \eta_c\gamma]$ (in keV) vs. $\mu$ (in GeV). 
Dashed line is the prediction in NR limit, solid line includes ${\cal O}(v^2)$ corrections, 
and  the band represents the measured width with error taken from \cite{Eidelman:2004wy}.}
\label{psi:to:etac:curve}   
\end{figure}

For a  weakly-coupled system, the dynamics is largely governed by the  
perturbative static potential, {\it i.e.}, $V_0\approx -C_F\alpha(\mu)/r$, where
the natural choice of $\mu$ is around the typical three-momentum scale.
One also needs to specify the quark mass  
in (\ref{M1:v2:correction}). 
A naive input is to use $\hat{m}$, half of the center-of-gravity ground state mass.
However, this simple procedure may induce an error of order $v^2$, which shouldn't
be neglected according to power counting.
A more consistent way is to choose the $1S$ mass $m$, 
which is defined implicitly through 
     $ \hat{m} =  m - \left \langle 10 \left | p^2 / 2m \right| 10 \right\rangle 
     \approx m (1- C_F^2 \alpha_s(\mu)^2 /8)$.

Many of  terms in \eqref{M1:v2:correction} are practically negligible.
Since  $\kappa_Q$ retains its NRQCD value and is only a few per cent for charm and bottom,
we may simply put it to be zero  in those ${\cal O}(v^2)$ terms (as a result, $I_3$ can be dropped
in both allowed and hindered transitions).
In fact, only $I_2$ in (\ref{M1:v2:correction}) accounts for genuine ${\cal O}(v^2)$ correction
for allowed transition, and 
we end up with a simple expression
\bqa
\Gamma[J/\psi\to \eta_c \gamma]   
&\approx &  { 4 \alpha_{\rm em} e_c^2 \over  3 \hat{m}_c^2} \, k_\gamma^3\, 
\left[ 1+ 2\kappa_c - {2 C_F^2 \alpha_s^2(\mu) \over 3}  \right]\,.
\label{allowed:final:weak:coupling}
\eqa
Fig.~\ref{psi:to:etac:curve} shows a comparison between this formula and the data.
It seems that \eqref{allowed:final:weak:coupling} is
compatible within the error to the data when $\mu$ is about $0.8$ GeV. Note this
value is consistent with the empirical $mv$ value for $J/\psi$. 
Therefore, our reasonable success in describing $J/\psi\to \eta_c\gamma $  may be viewed
as {\it a posteriori} support for the weak-coupling assignment.
However, rather sharp $\mu$ dependence may suggest that $J/\psi$ is not far from the
strong-coupling regime. 
One can employ \eqref{allowed:final:weak:coupling}  to  $\Upsilon\to \eta_b \gamma$  with more confidence,
and finds a smaller ${\cal O}(v^2)$ correction with flatter scale dependence. 
Unfortunately, very narrow width of $2\sim 3 \;\,{\rm eV}$  makes this transition unlikely to be detected.
 
For hindered transitions, $\it e.g.$ $\Upsilon^\prime(\Upsilon^{\prime\prime})\to \eta_b\gamma$,
the experimental upper bounds are already rather tight and  many model predictions have been
ruled out~\cite{Artuso:2004fp}. 
To be conservative, one would not expect \eqref{M1:v2:correction} to reliably describe 
these transitions if excited bottomonia are indeed in the strong-coupling regime. 
Nevertheless, it may not be too optimistic to treat, at least $\Upsilon^\prime$,
as being in weak-coupling regime. Proceeding along this line, 
we find that the $I_4$ term 
is dominating over $I_1$ and $I_2$, and the latter two nearly cancel each other. 
As a result, the predicted width is about an order-of-magnitude larger than the experimental upper bound! 
Even though there are lots of uncertainty associated with $I_4$,
this alarmingly large discrepancy seems to indicate that the weak-coupling 
 assignment of
$\Upsilon^\prime$  is problematic and a strong-coupling analysis might be more appropriate.

We end with a brief discussion on the $P$-wave $M1$ transitions, which have received less attention 
in the past, due to lack of phenomenological impetus.
Recent discovery of the $h_c$ state~\cite{Rosner:2005ry} may arouse the interest to study these processes.
Color-octet effect again vanishes because of the exactly same reason.
It is straightforward to apply the pNRQCD formalism to derive the  ${\cal O}(v^2)$ corrections.
For simplicity, we just quote the weak-coupling formula for the allowed transition: 
\bqa
\Gamma[n\:^3P_J \to n \:^1P_1 +\gamma] &=&  
{ 4 \, \alpha_{\rm em} \, e_Q^2 \over  3 m^2} \,k_\gamma^3\, 
\left|1+\kappa_Q - c_J\left \langle n1 \left| {{\bf p}^2 \over  m^2}\right|n 1\right \rangle  \right|^2 \,,
\label{M1:v2:correction:allowed:P-wave}
\eqa
where $c_J= 1/2, 1, 4/5$ for  $J= 0,1,2$.
For the $n\:^1P_1 \to n \:^3P_J \gamma$ transition, one simply multiplies the right side of 
(\ref{M1:v2:correction:allowed:P-wave})  by a statistical factor of $(2J+1)/3$.
It should be understood that this formula may be of limited use,
since $P$-wave onia may necessarily live in the strong-coupling regime.

The fine splittings between $h_c$ and $\chi_{c0}$, $\chi_{c1}$ and $\chi_{c2}$ are about 110, $-13$,
$-32$ MeV, respectively.  It seems that only $h_c\to\chi_{c0}\gamma$  has a serious chance to be observed,  
with a width comparable to $\Gamma[J/\psi\to \eta_c\gamma]$.
Future experimental input will enable us to infer the size of  relativistic corrections, thus
enriching our understanding of $P$-wave onia.


\begin{theacknowledgments}
 I thank N.~Brambilla and A.~Vairo for collaboration on this work, 
 also for their comments on the manuscript. 
 This research is supported in part by  Marie Curie ERG grant
under contract MERG-CT-2004-510967 and by INFN.
 
\end{theacknowledgments}



\bibliographystyle{aipproc}   

\bibliography{sample}

\IfFileExists{\jobname.bbl}{}
 {\typeout{}
  \typeout{******************************************}
  \typeout{** Please run "bibtex \jobname" to optain}
  \typeout{** the bibliography and then re-run LaTeX}
  \typeout{** twice to fix the references!}
  \typeout{******************************************}
  \typeout{}
 }

\end{document}